\documentclass{aa}  
\usepackage{hyperref}
\usepackage{graphicx}
\usepackage{natbib}
\usepackage{soul}
\usepackage{changes}

\usepackage{txfonts}
\usepackage{multicol}
\usepackage{multirow}
\usepackage{xcolor}
\newcommand{\e}{\mathrm{e}}

\begin{document}

   \title{Hadronic acceleration in the young star cluster NGC 6611 inside the M16 region unveiled by Fermi-LAT}

  \subtitle{Constraints on the acceleration efficiency}
%\titlerunning{M16}
   \author{G. Peron
          \inst{1}\thanks{\email{giada.peron@inaf.it}}
          \and
          S. Menchiari \inst{2,1} \and G. Morlino  \inst{1} \and E. Amato  \inst{1} 
          }

   \institute{INAF Osservatorio Astrofisico di Arcetri, Largo Enrico Fermi, 5, 50125, Florence, Italy
              \email{giada.peron@inaf.it}
         \and
             Instituto de Astrof\`{ı}sica de Andaluc\`{ı}a, CSIC, 18080 Granada, Spain}

   \date{Received ---; accepted ---}

% \abstract{}{}{}{}{} 
% 5 {} token are mandatory
 
  \abstract
   {Young Massive Star Clusters, long considered as potentially important sources of galactic cosmic rays, have recently emerged as gamma-ray emitters up to very high energies. } 
   {In order to quantify the contribution of this source class to the pool of Galactic CRs, we need to estimate the typical acceleration efficiency of these systems.}
   {We search for emission in the GeV band, as most of the energy is emitted in this band. We perform an analysis of Fermi-LAT data collected towards the M16 region, a star-forming region also known as the Eagle Nebula, which hosts the Young Massive Star Cluster NGC~6611. We model the acceleration at the stellar wind termination shock and the propagation through the wind-blown bubble to derive the energetics of the process and interpret the GeV observations.}
   {We find significant GeV emission in correspondence of a molecular cloud associated to the Young Massive Star Cluster NGC~6611. We interpret this as hadronic emission associated to particle accelerated at the cluster wind termination shock and propagated through the low-density wind-excavated bubble to the cloud. Our modeling allows us to put firm constraints on the acceleration efficiency in NGC~6611, assessing it between  $\sim 1$\% and $\sim 4$\%} %.}
   {}

%\titlerunning
\maketitle
 
\section{Introduction}
The list of sources responsible for the acceleration of  Galactic Cosmic Rays (GCRs) is currently under revision, both due to the fact that supernova remnants do not seem able to account for all the observed properties of CRs and to the discovery of new classes of gamma-ray emitters, including clusters of young massive stars. Star clusters (SCs) are promising CR contributors, because their energetics could account for a large fraction of the power retained in GCRs, and because they could naturally explain the anomaly in the chemical composition of CRs with respect to the interstellar medium (ISM) \citep{Tatischeff2021TheComposition, Gabici2023CosmicClusters}. 

The gamma-ray emission originating from SCs is often ambiguous in nature: some SCs seem to have dominant leptonic emission \cite[e.g. Westerlund~1][]{Aharonian2022A1,Haerer2023}), others are convincingly hadronic accelerators \citep{Peron2024ThePopulation}, and for a few objects the acceleration site and consequently the emission mechanism are still debated, like in the case of the Cygnus OB2 association \citep{Menchiari2024,Vieu2024}. Yet, the nature of the emission is a fundamental issue when evaluating the contribution of this source class to the pool of GCRs.

Here, we present the analysis of gamma-ray data obtained from the M16 complex, and we argue that the emission is predominantly hadronic. M16, also known as the Eagle Nebula, is one of the best-known H\textsc{ii} regions within a few kpc. It hosts, and is heated by, the young massive star cluster (YMSC) NGC~6611, which contains a total of 52 early type stars, among which a dozen are O stars. NGC~6611 has an estimated age of $\sim$1.3$\pm$0.3~Myr \citep{bonatto2006mass}.
Being so young, no supernova is believed to have exploded in the region yet, freeing the emission from contamination of these sources and their remnants. Notably, the cluster is also too young to host WR stars, which appear in a later evolutionary phase. This is also confirmed by available catalogs \cite[e.g.][and later updates]{Rosslowe2015} where no WR is associated to NGC~6611. At the same time, being the cluster still embedded in the gas cocoon from which it formed, we can find a dense gas layer just outside it, that serves as target for nuclear interactions, and the production of hadronic gamma-ray emission. This is common in young stellar clusters, but the case of M16 is special, in that an interacting giant molecular cloud, shaped by the wind of the stars, is unambiguously detected at the edge of the wind-blown bubble \citep{nishimura2021forest}. This allows us to model the acceleration and propagation of particles from the termination shock to the cloud, and to derive clear constraints on the efficiency of particle acceleration in the region.

The paper is structured as follows: in Sec.~\ref{sec:observations} we describe the gamma-ray and gas measurements towards the region; in Sec.~\ref{sec:modeling} we outline the model of acceleration and propagation applied to the system and derive constraints on the acceleration efficiency; finally, in Sec.~\ref{sec:conclusion} we summarize our findings and discuss their implications.

\section{Observations} \label{sec:observations}
\subsection{Fermi-LAT analysis}\label{subsec:fermi}
We carried out an analysis of Fermi-LAT data collected for more than 16.5 years (from August 2008 to March 2025) from the region of M16, and precisely centered at $(l_0,b_0)=(17,1)^\circ$. We applied standard data-quality cuts and standard models for background sources, as detailed in the Appendix \ref{app:spectrum}. In the region of M16, the unidentified source 4FGL~J1818.0-1339c is listed in the 4FGL catalog: we removed it from the source model and re-modeled the emission. The residual TS map is shown in Fig. \ref{fig:gas}.  We tested the position and extension of this source considering only events with energy $>$ 1 GeV, and obtained a best-fit position of $(l_*,b_*) = (17.10 \pm 0.05, 0.93 \pm 0.05)^\circ$ and a best-fit extension of $0.27^{\circ~+0.07^\circ}_{~-0.05^\circ}$, with a $TS_{ext}$ of 17.8 and a total detection significance of the source of 7.5~$\sigma$. 

We modeled the emission as a uniform disk with the reported extension and coordinates. Such a morphology is then used to extract the spectral points in the entire energy range (60 MeV--870 GeV). The resulting spectral points are shown in Fig. \ref{fig:sed}, and are derived {with a standard procedure, namely by fitting the normalization of the flux in each considered energy bin} \footnote{see \href{https://fermipy.readthedocs.io/en/latest/advanced/sed.html}{https://fermipy.readthedocs.io/en/latest/advanced/sed.html}}. %{The choice of the slope in each bin does not influence the final result, as described in Appendix.} 
%fitting \ea{the normalization ????} of a power-law of index 2 in each energy bin. 
%\begin{figure}
%    \centering
%    \includegraphics[width=1\linewidth]{ts_m16.pdf}
%    \caption{Test statistics (TS) map above 1 GeV in the region around M16. The position of the cluster NGC~6611 is indicated as a red star, and its relative termination shock and bubble radii are indicated as red and dotted-red circles. The light-blue and cyan contours indicate CO gas column densities of $2 \times 10^{22}$ and $2.5 \times 10^{22}$ cm$^{-2}$ and obtained in the velocity range 15-30~km~s$^{-1}$. The magenta circles indicate the position of Fermi sources from the 4FGL catalog \cite{Ballet2023} and their relative position uncertainties. }
 %   \label{fig:ts_map}
%\end{figure}

\subsection{Gas derivation}
\label{subsec:gas} 
\begin{figure*}
    \centering
  \includegraphics[width=0.33\linewidth]{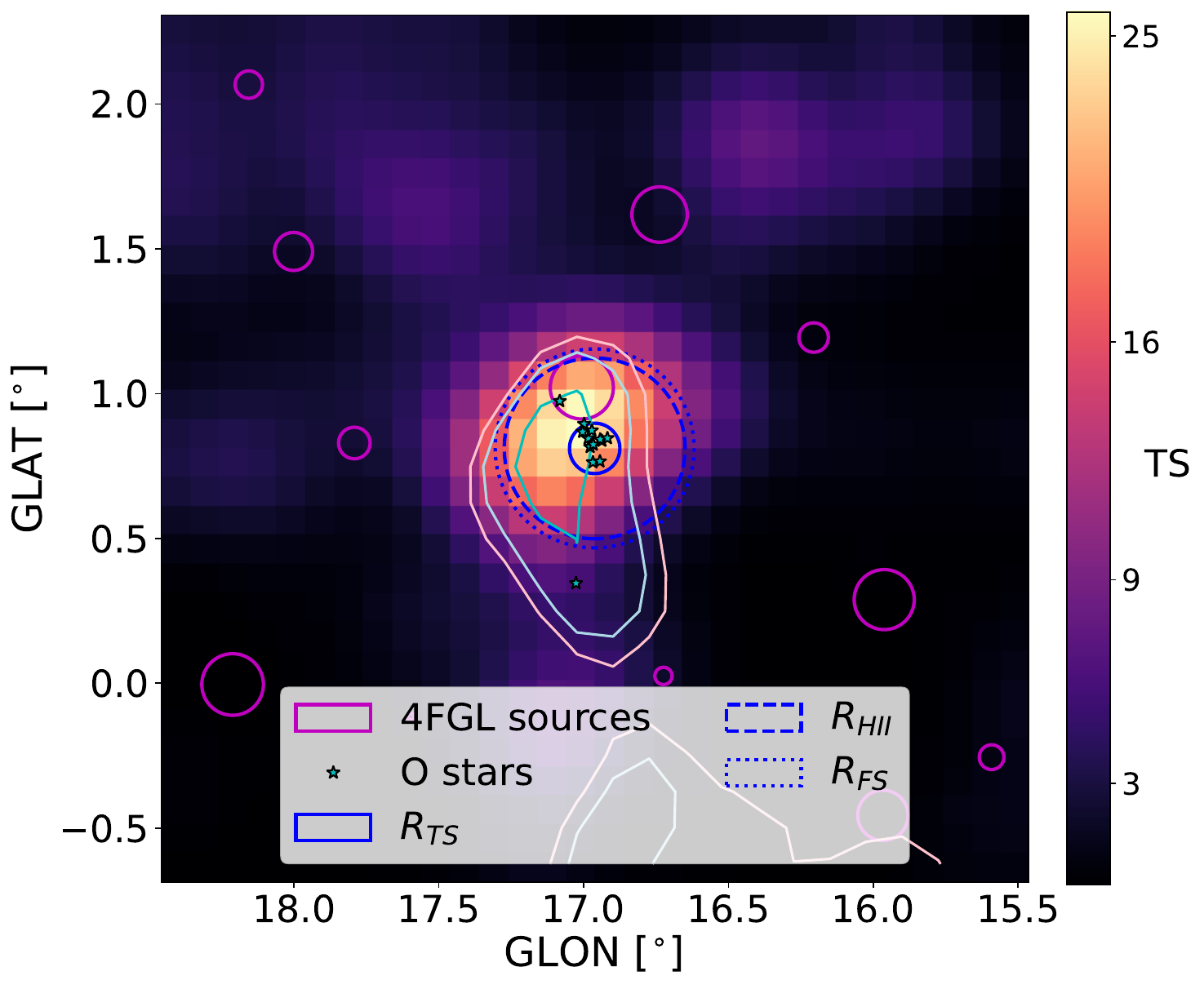}\includegraphics[width=0.345\linewidth]{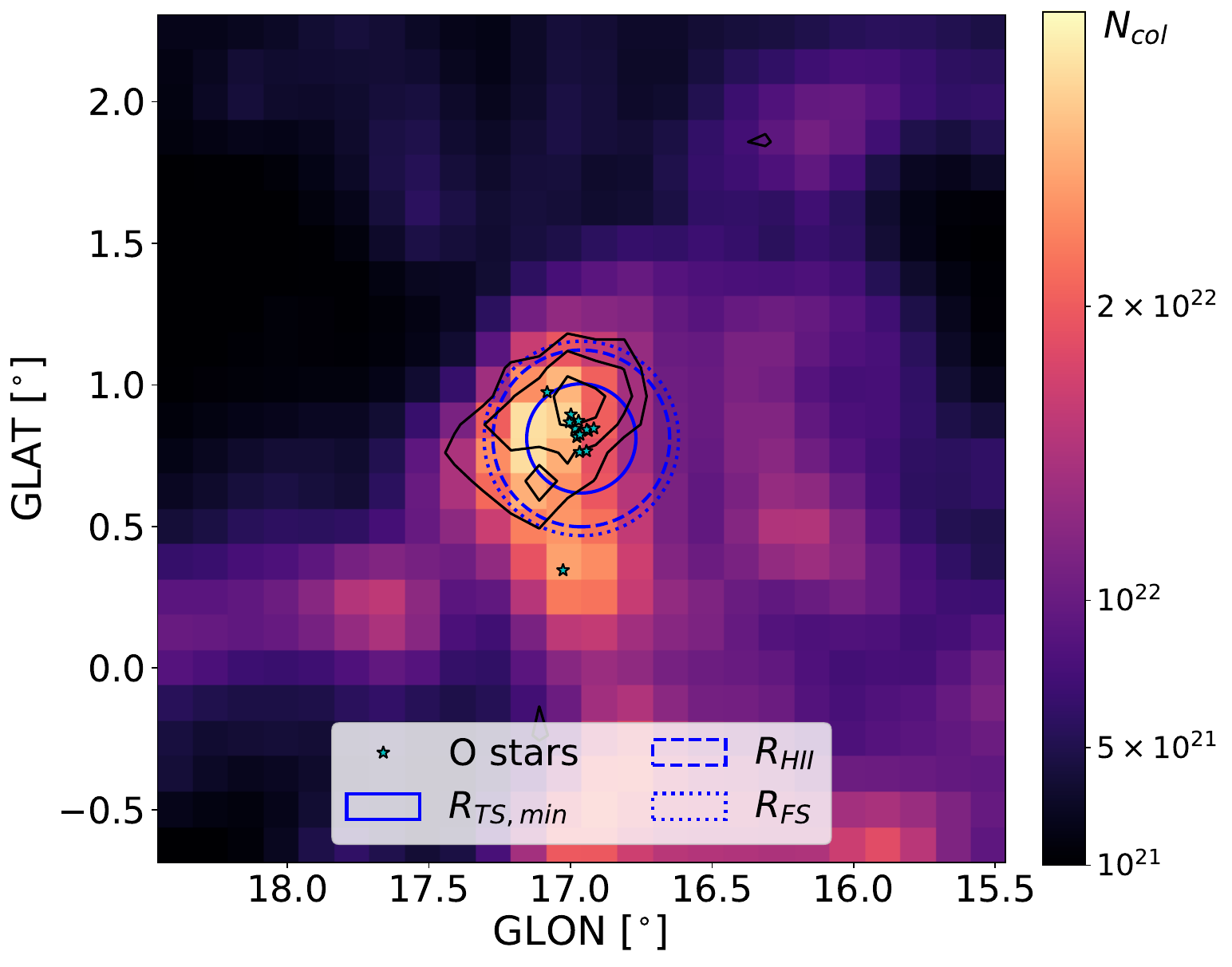}\includegraphics[width=0.325\linewidth]{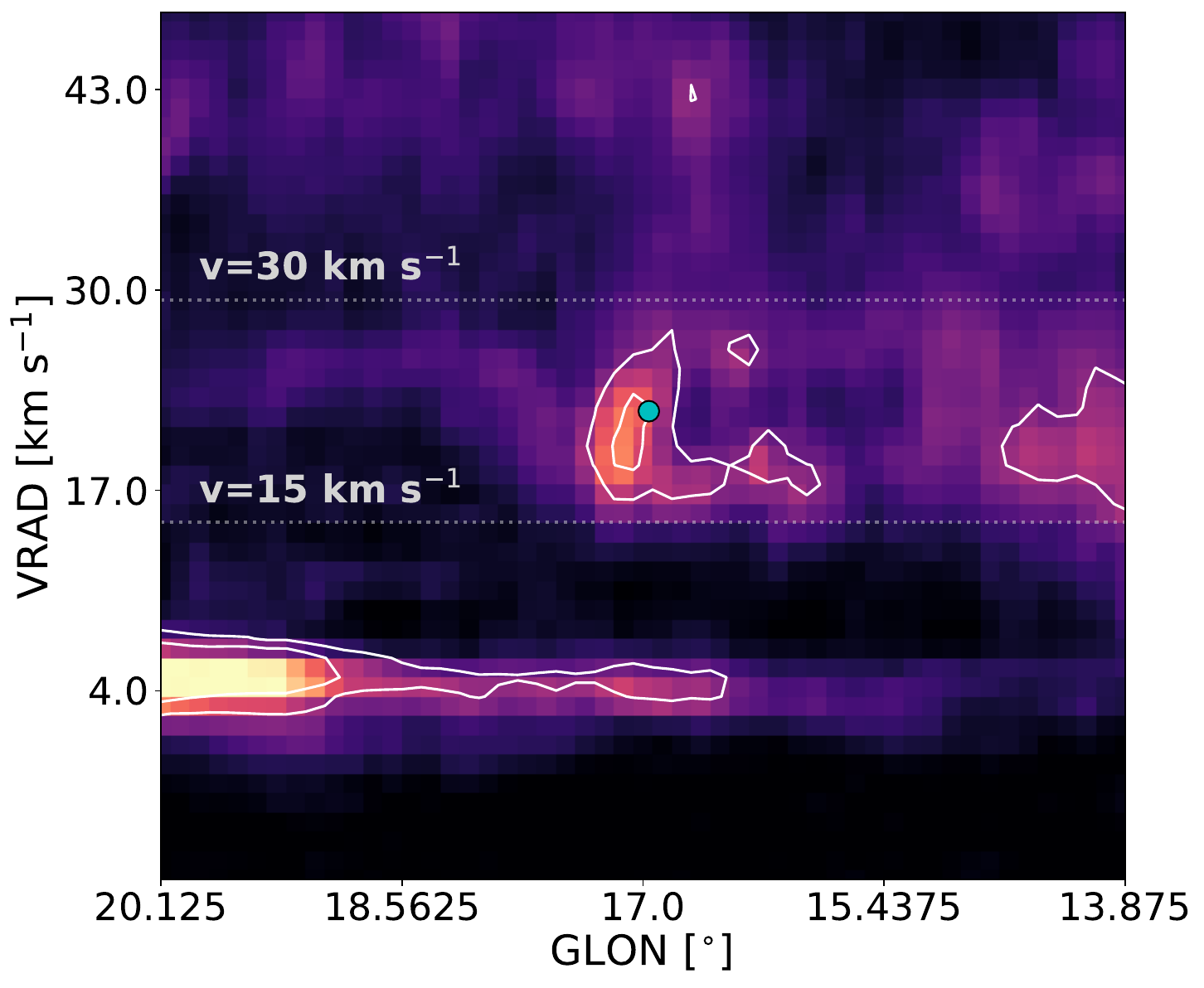}
    \caption{\textit{Left:} Test statistics (TS) map in the region around M16 derived from Fermi-LAT observations. The position of the O stars in the cluster NGC~6611 are indicated as blue stars, and its termination and forward shock radii are indicated a solid and dotted blue circles, while a dashed circle indicates the limit of the H\textsc{ii} region. The light-blue, cyan, and pink contours indicate CO gas column densities of $1.5 \times 10^{22}$, $1.7 \times 10^{22}$ and $2.3 \times 10^{22}$ cm$^{-2}$, obtained in the velocity range 15-30~km~s$^{-1}$. The magenta circles indicate the position of Fermi sources from the 4FGL catalog \cite{Ballet2023} and their relative position uncertainties, evaluated as their 68\% confidence radius. \textit{Center:} molecular gas distribution in the region, as traced by CO line emission, integrated over the line of sight in the velocity range 15 km/s, 30 km/s. The black contours refer to Fermi-LAT test-statistics and represent 3,4, and 5 sigma levels. \textit{Right:} the velocity-longitude distribution of CO. The map is integrated over the whole latitude range. The star marker indicates the position of NGC~6611.}
    \label{fig:gas}
\end{figure*}

{To derive the gas distribution}, we analyzed the CO(J=2$\rightarrow$1) line emission map provided by \cite{Dame2000}, which traces the dense molecular gas. {Results are} shown in Figure \ref{fig:gas} (central and right panel) both in the longitude-velocity and in the longitude-latitude planes. We see that at the selected coordinates, there is an isolated gas structure located within the radial velocity range 15-30~km~s$^{-1}$. This well known structure corresponds to a molecular cloud associated to the cluster, as it appears to be compressed by the interaction with it \citep{nishimura2021forest}. The gamma-ray emission (shown in the left panel of Fig.~\ref{fig:gas}) is spatially coincident with this gas structure, suggesting that the emission actually originates from the interaction of high-energy nuclei with this cloud. 
From the CO map we estimate the gas density associated to this system, which is used both to compute the gamma-ray emissivity, and to estimate the size of the bubble created by the stellar wind, as discussed in Sec. \ref{sec:modeling}. 
We assume a conversion factor from CO to H$_2$ of $X_{\rm CO}= 2 \times 10^{20} \rm{cm^{-2}~K~km~s^{-1}}$ \citep{Bolatto2013}, and use the all-sky map of HI line emission by \citet{BenBekhti2016} for the atomic component, with a conversion factor of $X_{\rm HI}= 1.8 \times 10^{18} \rm{cm^{-2}~K~km~s^{-1}}$. Instead, the mass of the ionized component is neglected, because it can be demonstrated that this is much more diluted compared to the neutral component. 
We use the radius of the H\textsc{ii} region, as traced by WISE \citep{Anderson2014TheRegions}, as the extension of the ionized bubble. A dense layer of swept-up gas is assumed to occupy a shell in the immediate surroundings of the bubble. We assume that the radius of the shell is $\sim$5\% $R_{HII}$ which is consistent with the values obtained is simulations \citep{Gupta2016}, therefore we measure the mass contained within 1.05 $R_{\rm HII}$. Taking the distance measured by Gaia of 1646 pc\citep{Cantat-Gaudin2020}, the resulting mass of the swept up material is 23.6~$\rm kM_{\odot}$. 
%Mass radial distribution 

\section{Derivation of acceleration efficiency}\label{sec:modeling}
We assume the collective wind termination shock from the cluster NGC~6611, as the primary site for particle acceleration, and apply the model derived by \citet{Morlino2021} to describe the emission and infer the acceleration efficiency. 
%Luminosity
The wind luminosity of NGC~6611 has been estimated by \citet{Celli2024} to be $L_{W,1}=6.3 \times 10^{36}~\rm{erg~s^{-1}}$, which corresponds to a total stellar mass $M_{*,1} = 1697~\rm{M_{\odot}} $. However, the latter is considered by the same authors to be a strict lower limit, because of the limitation of their method in resolving the stellar population in young clusters. A similar value for the cluster mass is derived independently by \citet{Bonatto2006}, who also argue that the value should be considered as a lower limit, because it results from the distribution of stars less massive than 5~M$_\odot$. A later study \citep{Selim2016} derived a mass of 2260 M$_{\odot}$ based on observations of stars up to 35 M$_\odot$, while estimates by \citet{Stoop2023} suggest a mass range between 5 and 10 $\rm{kM}_{\odot}$, when extrapolating the mass distribution up to stars 100 M$_{\odot}$. We therefore considered a second value for the wind luminosity, $L_{W,2}=3.8 \times 10^{37}~\rm{erg~s^{-1}}$, which we derived following the method presented by \citet{Menchiari2025}: we simulated the stellar population of the cluster, assuming the initial mass function reported by \cite{Kroupa2001} and fixing the total mass of the cluster to $M_{*,2}=3~M_{*,1}$\footnote{The choice of $M_{*,2}=3~M_{*,1}$ is motivated by the estimation of the systematic uncertainty that \citet{Celli2024} argue to affect their sample. Moreover, the estimated value is consistent with the mass range derived by \citet{Stoop2023}}. 

\begin{figure}
    \centering
    \includegraphics[width=1.\linewidth]{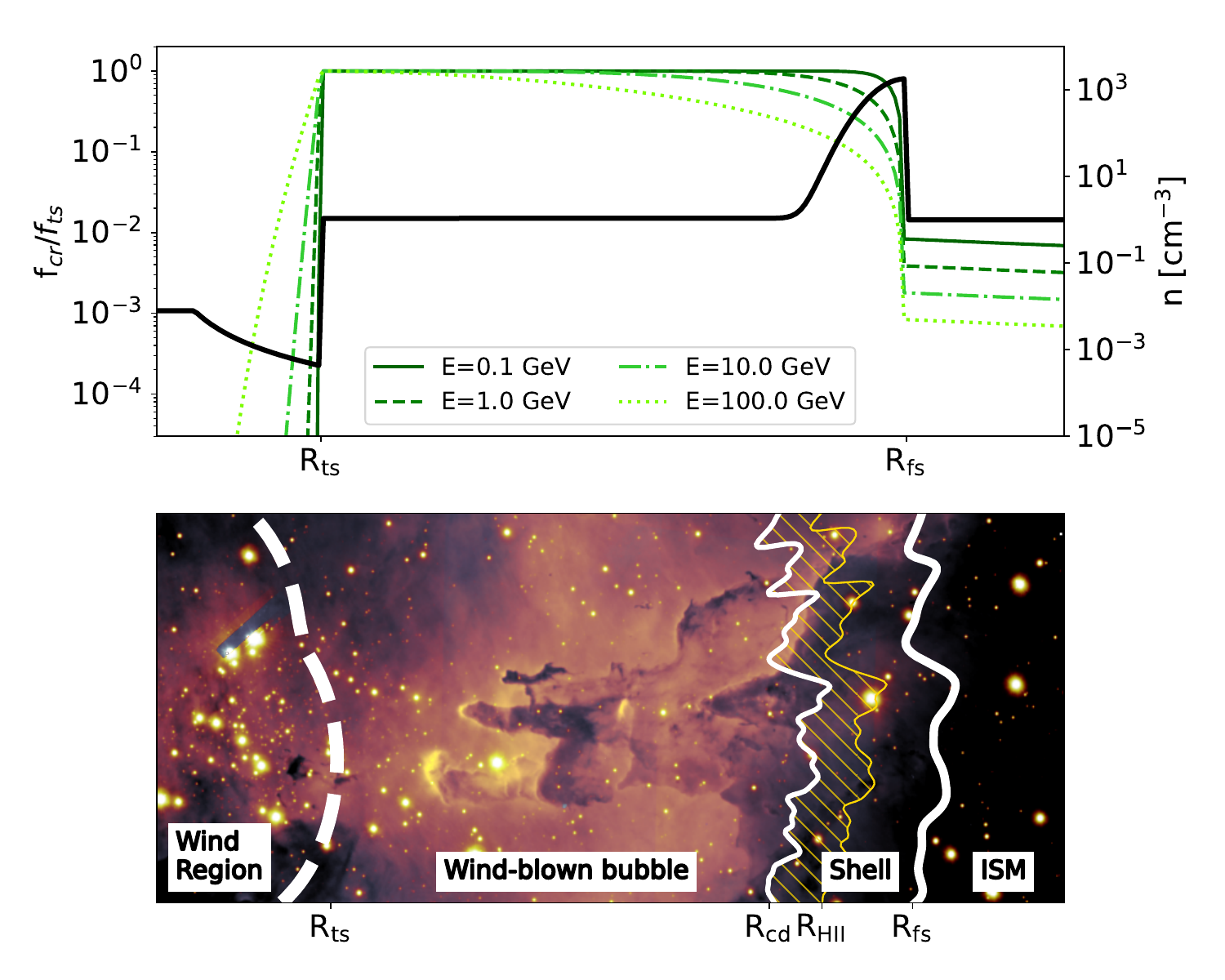}
    \caption{Representation of the geometry of the system: in the upper panel, the distribution of gas density as a function of the radial distance from the center of the cluster (black curve), and the radial distribution of accelerated particles at different energies (green solid, dashed, dashed-dotted and dotted lines), considering propagation in Kraichnan-like turbulence. In both cases the highest value for the wind luminosity, $L_{w,2}$, is assumed. In the lower panel, a sketch of the geometry of the system projected on the sky is superimposed to a cutout of an image of the system obtained in otpical and through an H$\alpha$ filter (VPHAS+ survey%[https://ui.adsabs.harvard.edu/abs/2014MNRAS.440.2036D/abstract]
    ). }
    \label{fig:geometry}
\end{figure}

%Magnetic field and turbulence
The wind luminosity powers the acceleration and injects turbulence into the system; therefore, it contributes to efficient confinement of the accelerated particles.  To compute the diffusion coefficient, we assumed that a fraction $\eta_{B}=5\%$ of the wind power is converted into magnetic turbulence, with a coherence length of $l_c=1~\rm{pc}$, and considered the three turbulence spectra most commonly adopted in astrophysics: Bohm, Kraichnan, and Kolmogorov. The magnetic field that results from this hypothesis is $4.7~\mu$G at the wind-termination shock, which corresponds to a value of $\sim 13~\mu$G downstream. The latter is largely below the upper limit found by \cite{Mackey2011} based on magneto-hydrodynamical simulations of the famous pillar structures (the {\it pillar of creation}) observed in M16. 
%Shell shape
We then modeled the geometry of the system as sketched in Fig. \ref{fig:geometry}.  We considered that most of the mass that we measure from the gas tracers is swept up to form a dense shell, whereas the mass inside the bubble is mostly contributed by the evaporation of the shell \citep{Castor1975}, and the innermost parts of the bubble ($r<R_{\rm TS}$) are populated by the stellar wind material, computed from $\dot{M}$, which is in turn derived from the cluster mass as in \cite{Menchiari2025}. In a 1D model the shell thickness is very small, as a consequence of radiative cooling. However, HD instabilities fragment the shell, increasing its effective thickness. According to \cite{El-Badry2019}, the typical length-scale to which the fragmentation instability grows in a time $\Delta t$ is of the order of $\lambda \sim 2 \pi \Delta t~v_{\rm rel}\,r_{\rho}^{-1/2}$
%To estimate such an effect, we followed the prescription by \cite{El-Badry2019}, who \ea{estimate the typical length-scale to which the fragmentation instability grows in a time \st{consider the growth of instabilities in a time} $\Delta t$} to be of the order of}{At the same time, a contribution of mass inside the bubble could originate from fragmentation of the shell itself. To evaluate the impact of this effect, we estimate the typical scale at which HD instabilities form during the age of the cluster, following the prescription presented in El-Badry2019 who consider the scale of the perturbation of the shell to be of the order of} 
where $v_{\rm rel}$ is the shear speed between the shell and the bubble while $r_\rho = \bigg(\frac{\rho_{\rm high}}{\rho_{\rm low}} \bigg)$ is their density ratio. 
By fixing $\Delta t$ to the cluster age and considering reasonable estimation for $v_{\rm rel}~\sim 10~\rm{km~s^{-1}}$ , and {$r_{\rho}\approx 10^3$, as inferred from} the observed densities, the relevant scale of instabilities turns out to be of the order of 2\,pc. 
{In order account for this, we assumed the mass of the shell to be distributed according to $\exp[-((R_{cd}-r)/0.5 {\rm pc})^2]\ \theta(R_{cd}-r)\ \theta(R_{cd}-r+2 {\rm pc})$, where $\theta$ is the Heavyside function, and the last term takes into account the above estimate of the penetration length.}
We {neglect, instead,} the mass of larger but more rare structures like the pillars;  their estimated mass is $\sim 200~\rm{M_\odot}$ \citep{McLeod2015}, which amounts to only  $\sim 1 \%$  of the swept-up mass of the shell.  
%Rfs and mechanical efficiency
Finally, we constrain the size of the wind-blown bubble, namely the location of the forward-shock, $R_{\rm fs}$, based on the observed size of the H\textsc{ii} region. The latter is 0.312$^{\circ}$ \citep{Anderson2014TheRegions}, yielding $R_{\rm fs}=1.1~R_{\rm HII}= 9.8~{\rm pc}$, at the given distance. By comparing the observed radius, with what is expected from Weaver's equation, modified to take into account losses:
\begin{equation}
R_{\rm fs} =  18.14 ~\bigg(\frac{\eta_m L_w }{10^{37 }\rm{erg~s^{-1}}} \bigg)^{\frac{1}{5}} \bigg(\frac{n_0}{197~\rm{cm^{-3}}} \bigg)^{-\frac{1}{5}} \bigg(\frac{t}{1.3~\rm{Myr}} \bigg)^{\frac{3}{5}} ~\rm{pc}, 
\end{equation} we can derive the mechanical efficiency, $\eta_m$, which accounts for the fraction of wind-luminosity that goes into inflating the bubble \cite[see, e.g.][]{Vieu2022CosmicSuperbubbles}. The density $n_0$ is computed from the gas map described in Sec. \ref{subsec:gas} , over the volume of the bubble, which gives $n_0=197~\rm{cm^{-3}}$. The resulting value of $\eta_m$ is then $\sim 8\%$ for $L_{w,1}$ and  $\sim 1.3\%$ for $L_{w,2}$,in good agreement with the values found in the simulations by \citet{Yadav2017HowSuperbubbles} for young systems in a dense medium. The value of $\eta_m$ also defines the size of the termination shock, resulting in  $R_{\rm TS}= 2.17$~pc and  $R_{\rm TS}= 5.6$~pc for the two values of the luminosity (see Eq. \ref{eq:Rts}). 

\begin{figure}
    \centering
    \includegraphics[width=1\linewidth]{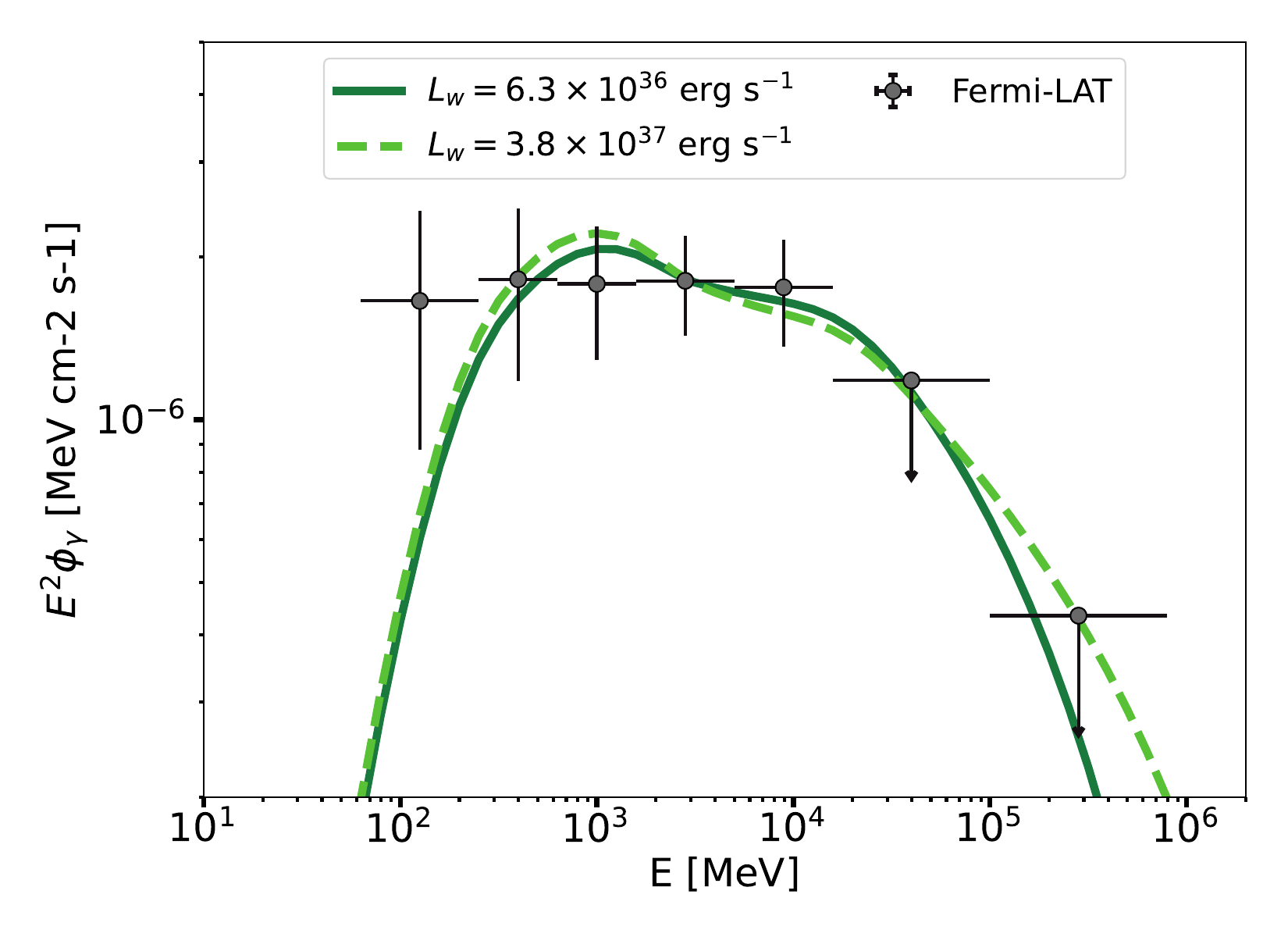}
    \caption{Spectral energy distribution (SED) of the source detected by Fermi-LAT in correspondence of NGC~6611/M16. The solid and dashed green lines represent the model of emission in the case of Kraichnan turbulence, and differ in the assumed wind luminosity, $L_w$, as indicated in the figure legend.}
    \label{fig:sed}
\end{figure}

Once the wind luminosity, the propagation regime, and the density profile are fixed, the gamma-ray emission depends only on the CR distribution, which we assume to follow the solution of \cite{Morlino2021}, namely a power-law in momentum with a cutoff that depends on the turbulence spectrum $\propto p^{-\alpha_{\rm CR}} e^{\Gamma(p)}$ (see extended functional form in Appendix \ref{app:acc_prop}), and on the acceleration efficiency, $\eta_{CR}$, defined as the fraction of $L_{w}$ that goes into accelerated particles. For each considered model of diffusive transport (Kolmogorov, Kraichnan, and Bohm), we derive the best combination of $\alpha_{\rm CR}$ and $\eta_{\rm CR}$ that fits the gamma-ray spectral energy distribution measured with Fermi-LAT. The goodness of the fit is evaluated with a $\chi^2$-test, which is computed over a grid of possible values of $\alpha_{\rm CR}$ and $\eta_{\rm CR}$. More details on this procedure are provided in Appendix \ref{app:fit}. An example of the best-fit SED is shown in Fig.~\ref{fig:sed}, together with the measured spectral points. The calculation is for a Kraichnan-type diffusion. The plot makes it clear that the assumption we made for the luminosity influences the spectral shape only at the highest energies, where unfortunately we do not have significant detection. The results for the different cases are reported in Table \ref{tab:results}, where we report the best fit values for acceleration efficiency and spectral slope, together with the corresponding $\chi^2$.

\begin{table*}[]
\centering
\small
\begin{tabular}{|ccc|ccc|ccc|ccc|ccc|}
\hline
    {$L_{\rm w}$}  & {$\dot{M}$} &   $\eta_{\rm m}$   &\multicolumn{3}{|c|}{$E_{\max}$ [TeV]}&\multicolumn{3}{c}{$\eta_{\rm CR}$} &  \multicolumn{3}{|c|}{$\alpha_{\rm CR}$} & \multicolumn{3}{c|}{$\chi^2$}   \\  
     %\hline
       erg ~s$^{-1}$ &  M$_\odot$ yr$^{-1}$& &  Kolm. & Kra. & Bohm &Kolm. & Kra. & Bohm &  Kolm. & Kra. & Bohm & Kolm. & Kra. & Bohm \\
     \hline
 6.3 $\times 10^{36}$ & 2$\times 10^{-6}$ & 8\%&1.4& 20.6&299.4&0.417& 0.145& 0.012& 2.46& 3.74& 4.2& 1.64& 0.77& 0.77\\
%& 6.3e+36 & 2e-06 & 0.01 & 0.166 & 0.058 & 0.013 & 2.14 & 3.8 & 4.24 &1.78 &0.81 &0.97 \\
3.8 $\times10^{37}$  & 1.4$\times 10^{-5}$&  1.3\%   &18.8& 115.3&708.5&0.091& 0.036& 0.012& 3.6& 3.92& 4.26& 0.95& 0.87& 1.16\\
%3.8e+37 & 1.4e-05 & 0.01 & 0.044 & 0.02 & 0.013 & 3.3 & 4.02 & 4.26 & 1.19 & 0.83 &1.3  \\
\hline
\end{tabular}
    \caption{Summary of the parameters that characterize the system: the luminosity, $L_w$, and mass-loss rate, $\dot{M}$, of the cluster wind, the mechanical efficiency, $\eta_m$. For each considered turbulence scenario  and for both values of the luminosity, we also report the result of the best fit to the Fermi data-points, namely the maximum energy $E_{\max}$, the acceleration efficiency, $\eta_{\rm CR}$, and the injection spectral index, $\alpha_{\rm CR}$, along with the corresponding $\chi^2$. }
    \label{tab:results}
\end{table*}

\section{Discussion and conclusions} \label{sec:conclusion}
We have detected significant gamma-ray emission from the M16 region in the energy range from 60 MeV to 20 GeV. The emission, compatible with an extended source of radius $\sim 0.3^{\circ}$, is found in the vicinity of the young star cluster NGC~6611, in coincidence with the location of a gas cloud that interacts with the wind-blown bubble of the system. The spatial coincidence with dense target gas suggests a hadronic origin of the emission, which we interpret as due to particles accelerated at the wind termination shock of the cluster {and then propagating to the cloud where the interaction takes place.}
%\replaced{that propagate towards the cloud where the interaction take place}{, and then reach the cloud by propagating through the turbulent bubble}. 
We considered a non-uniform mass distribution {inside} the bubble. This affects the resulting gamma-ray emission, as the particles that are accelerated at $R_{\rm TS}$ need to propagate and reach the shell before interacting to produce detectable gamma radiation. In the process, the CR distribution is shaped by propagation in the turbulent medium, which we account for when calculating the efficiency.  We also tested a possible contribution of leptonic origin and found that, with the assumed density, leptonic emission can be appreciable only if the electron-to-proton ratio of the accelerated particle population
%the ratio of electron-to-protons accelerated at the shock 
largely exceeds 0.01. For lower ratios, the pion-decay emission is dominant. As a consequence, the derived efficiency is solid within a factor $<$2 (see Appendix \ref{app:lepton}). 

%In such a scenario, we fitted the model of acceleration and propagation . The result is compatible within the probed energy range, with an acceleration at the shock with an injection spectrum of $p^{-\alpha_{CR}}$ and maximum energy. 
These considerations allowed us to derive the acceleration efficiency of the system, $\eta_{\rm CR}$.  We presented the results using upper and lower bounds for the wind kinetic luminosity derived from corresponding estimates of the cluster mass from the literature, but we consider the upper bound ($L_{w,2}$) closer to the real value, as the lower value derives from an underestimation of the mass. However, we also notice that the gamma-ray spectrum can be explained also using the lower bound and still reasonable values for the other parameters (see Table~\ref{tab:results}) 
The physics ingredient to which the fit parameters are most sensitive is the turbulence spectrum. We note that, assuming Kolmogorov type diffusion, fitting the data requires a very hard spectrum of accelerated particles, with $\alpha_{\rm CR} < 4$, and an acceleration efficiency between 10\% and 40\%, which is difficult to reconcile with DSA theory.
%We note that the fit with a Kolmogorov \replaced{diffusion}{profile} needs a very hard \ea{\st{acceleration}} spectrum \ea{of accelerated particles}, with $\alpha_{\rm CR} < 4$, \added{and an acceleration efficiency \ea{between within 10\% and 40\%,} which is difficult to reconcile with DSA theory. \deleted{As side effect, the acceleration efficiency needed  to reproduce the data is quite high, with values within 10\% and 40\%, which we believe are unreasonable.}  
The Bohm assumption, instead, requires a rather steep acceleration spectrum ($\alpha_{\rm CR}\sim 4.3$), as any harder spectral shape would violate the upper limits from observations (see also Appendix \ref{app:fit}). While we cannot exclude this injection spectrum, we remark that Bohm is generally considered an extreme case for turbulence, which provides an extremely efficient confinement. In a system with Bohm-like turbulence, an efficiency of 1.2\% would be sufficient to explain the observations. For the considerations above, the latter can be regarded as a stringent lower limit to the acceleration efficiency in the system. Finally, the Kraichnan case falls somewhat in between, with a slope $\alpha_{\rm CR}\sim 4$ and an efficiency of 3.6\%.
Our conclusion is that the acceleration efficiency can be constrained between 1.2\% and 3.6\%.
Interestingly, this is similar to the values derived by \cite{Peron2024ThePopulation} for the young embedded clusters RCW~32, RCW~36 and RCW~38, in the Vela molecular ridge, although in those cases the estimates were obtained under much rougher assumptions on particle confinement inside the wind bubble. 
Following the same approach proposed there, we can estimate the contribution of stellar winds to the production of CRs, assuming that all stellar wind systems accelerate, on average, with the same efficiency derived for NGC~6611. To do that, we consider that the total wind luminosity provided by OB stars over the Galaxy is estimated to be $1\times 10^{41} \rm{erg~s^{-1}}$ \cite{Seo2002}, and that the total luminosity of cosmic rays is $7\times 10^{40} \rm{erg~s^{-1}}$ \cite{Strong2010GlobalWay}.  With an efficiency of 3.6\% (1.2\%), the fraction of CR luminosity contributed by the winds of SC, $\epsilon_w = \eta_{\rm CR} L_{\rm tot,SC} / L_{\rm tot,CR}$, is 5.1\% (1.7\%). This value is well in agreement with the estimates derived by \cite{Tatischeff2021TheComposition} from consideration on the chemical composition of CRs, which requires a fraction $\epsilon_w\sim 6\%$ of CR contributed by stellar winds. Following \citet{Peron2024ThePopulation}, we can also derive the expected  ratio  $X_{Ne}=^{22}$Ne/$^{20}$Ne in CRs: in fact the relative abundance of $^{22}$Ne in the winds of massive stars is much larger than that in the ISM, with $X_{Ne,w}=1.56$ vs $X_{Ne,ISM}=0.0735$. Considering the derived values of $\epsilon_w$,  $X_{Ne, \rm CR} = \epsilon_w X_{Ne,w} +(1-\epsilon_w) X_{Ne, ISM} = 0.148~(0.099)$ . The latter can be compared with the measured value, $X^0_{Ne,\rm CR}=0.317 \sim 2 X_{Ne, \rm CR}$.  Despite the uncertainties, our value well approaches the measured one even if we neglected the role of Wolf-Rayet stars, believed to be the main contributors of $^{22}$Ne. More observations are needed to understand whether there is an evolution of the efficiency with the evolutionary phase of the cluster. 

As a final consideration, we evaluate the grammage at the source, namely the amount of matter traversed by CRs inside the cluster bubble before being released into the ISM. This issue was recently discussed by \cite{Blasi2025} in the context of SCs, showing that for the case of Cygnus OB2 and Westerlund~1, the grammage accumulated by the accelerated particles before leaving the cluster exceeds the grammage measured at Earth, implying that young SCs cannot be the main contributors of CRs.  We repeated the same calculation considering the values reported in Table \ref{tab:results} for NGC~6611, and the measured gamma-ray flux at 10 GeV. Our result shows that the grammage accumulated at the source is a factor 0.4 (0.2) of the measured grammage at Earth at 10 GeV ($\sim$7.5~g~cm$^{-2}$) for the Kraichnan (Bohm) case. However, considering that SCs contribute only a fraction $\epsilon_W$ of the total Galactic CRs, this has a very minor impact on the overall grammage accumulated by CRs.
Even considering the extreme scenario with the lower luminosity value of $L_{w,1}$, for which we obtain $\epsilon_w=0.2$, we find the ratio between grammage in the source and total of 0.3. Such a result suggests that acceleration at stellar winds does not significantly affects the ratio of primary over secondaries measured at Earth, at least at the energies probed here.  
\bibliographystyle{aa}
\bibliography{biblio}

\appendix

\section{Fermi-LAT analysis details} \label{app:spectrum}
For the Fermi-LAT analysis, we considered data in a region of interest (RoI) of 8$^{\circ}$ around $(l_0,b_0)= (17,1)^\circ$. We performed a binned analysis, considering  spatial  steps of 0.1$^\circ \times 0.1^\circ$, and 8 energy bins per decade, logarithmically spaced in the range between 60 MeV and 870 GeV. We applied the recommended data-quality cuts for source analysis (DATA\_QUAL$>$0 \&\& LAT\_CONFIG==1, with maximum zenith angle set to 90$^{\circ}$) and selected only events of type \texttt{evclass=128} converted both at the front and at the back of the detector (\texttt{evtype=3}). As a starting model for the emission, we used the latest released catalog of sources (4FGL-DR4,\citet{Ballet2023,Abdollahi2022IncrementalCatalog}), together with the Galactic and extra-galactic emission templates provided by the Fermi collaboration (namely \texttt{gll\_iem\_v07.fits} and \texttt{iso\_P8R3\_SOURCE\_V3\_v1.txt}). After a first optimization of the parameters, we included two additional point-like sources to account for two hot-spot in the residual test-statistic (TS) maps that exceeded a TS value of 9. The additional sources, located at $(l_1,b_1)=(17.933,-2.75)^\circ$ and $(l_2,b_2)=(17.949,-2.02)^\circ$ are well separated ($>1^{\circ}$) from the source of interest, and hence are of no concern in terms of systematic uncertainties. We performed an extension test, which positively detected an extension of 0.27$^{+0.07}_{-0.05}$ at a significance level of $TS_{ext} =17.5$. We found no appreciable difference in modeling the source as a Gaussian or as a disk, since the log-likelihood of the overall fit in the two cases differs by less than 0.1$\sigma$. We further tested a possible curvature in the spectral energy distribution, by comparing the likelihood of a power-law model with the likelihood of a log-parabola, and calculated $TS_{\rm curv} = 2(\log L_{\rm LP}-\log L_{\rm PL})$. The resulting $TS_{\rm curv}=8.37$ is well above the threshold ($TS_{\rm curv}>4$) set in the 4FGL catalog \cite{Abdollahi2022IncrementalCatalog} for preferring a curved spectrum over a power-law spectrum. This means that in the 4FGL catalog our source would be flagged as a log-parabola, even if the curvature significance is below 5, as  $ \sigma_{\rm curv} = \sqrt{TS_{\rm curv}} \lesssim 3$ . This relatively low threshold is adopted in the catalog because the power-law spectrum tends to overestimate the flux at the edge of the  sensitivity range,  and this is particularly important when using the catalog sources as background. Given that our aim is to characterize the flux using a more sophisticated modeling, based on acceleration and interaction of protons, we find the power-law description satisfactory for the overall fit. The resulting best fit parameters for the power-law are: $\alpha=2.14\pm 0.06$ and 
$F_0(E_0= 1~\rm{GeV})= (1.7 \pm 0.2) \times 10^{-12} ~\rm{MeV^{-1}~cm^{-2}~s^{-1}}$.
We report, for completeness, also the best-fit parameters obtained with a log parabola function, namely:
$$
F(E)=F_0\bigg( \frac{E}{E_b} \bigg)^{(-\alpha +\beta \ln (\frac{E}{E_b}))}
$$
with $F_0(E_0=1~\rm{GeV}) = (1.8 \pm 0.3) \times 10^{-12}~\rm{MeV^{-1}\,cm^{-2}\,s^{-1}}$,$\alpha = 1.77\pm 0.13$ and $\beta=0.151049\pm 0.000005$, while $E_b$ is fixed to 1 ~GeV.  

As a last step, we extracted the spectral energy distribution, by fitting in every energy bin the normalization of a power-law spectrum with spectral index 2. We checked that this choice does not affect the spectral points, by using the locally measured index, namely the spectral index derived from the overall fit, for each pixel. We further repeated the procedure with both the power-law and the logParabola models and by varying the size of the energy bins, finding no appreciable difference in the results.

%0.2657 + 0.0687 - 0.0542
%2025-04-16 11:15:47 INFO    GTAnalysis._extension(): TS_ext:        17.821
%2025-04-16 11:15:47 INFO    GTAnalysis._extension(): Extension UL: 0.3806
%2025-04-16 11:15:47 INFO    GTAnalysis._extension(): LogLike:  1239891.549 DeltaLogLike:       13.871
%2025-04-16 11:15:47 INFO    GTAnalysis._extension(): Position:
%(  ra, dec) = (  274.6458 +/-   0.0543,  -13.6079 +/-   0.0449)
%(glon,glat) = (   17.1043 +/-   0.0516,    0.9338 +/-   0.0480)

%gta.add_source('Src1',{'glon':17.949,'glat': -2.025  ,'SpatialModel':'PointSource','SpectrumType':'PowerLaw'} )
%gta.free_sources(free=False)
%gta.free_source('Src1')
%gta.fit()

%gta.add_source('Src2',{'glon':17.9310753,'glat':-2.7542902,'SpatialModel':'PointSource','SpectrumType':'PowerLaw'} )
%gta.free_sources(free=False)
%%gta.free_source('Src2')
%gta.fit()

\section{Modeling the acceleration and propagation within the wind-blown bubble} \label{app:acc_prop}

The model for particle acceleration and propagation adopted here is the one proposed by \cite{Morlino2021} and applied by \cite{Menchiari2024} to the Cygnus region. The model assumes that particles are accelerated at the wind termination shock and then propagate by advection and diffusion in the low-density bubble, made turbulent by the development of MHD instabilities.
We refer to the literature cited above for a full description of the solution. Here, we just recall the basic ingredients of the model. 
The termination shock radius is calculated as: 
\begin{equation}
\label{eq:Rts}
\begin{split}
R_{\rm ts} =26~ \eta_m^{-\frac{1}{5}} & \bigg(\frac{ \dot{M} }{10^{-4}~ \mathrm{M_{\odot} yr^{-21}}}\bigg)^{\frac{3}{10}}  \bigg(\frac{V_w}{2000 \mathrm{~km~s^{-1}}}\bigg)^{\frac{1}{10}}   \times \\ & \times \bigg(\frac{n_0}{10~\mathrm{cm^{-3}}}\bigg)^{-\frac{3}{10}} \bigg( \frac{t}{10~\mathrm{Myr}}\bigg)^{\frac{2}{5}}~ \mathrm{pc}     
\end{split}
\end{equation}.

The distribution function is obtained by solving the steady state transport equation in radial symmetry using two boundary conditions: net zero flux at the center of the bubble, matching of the Galactic CR distribution $f_{\rm gal}$ \cite[tuned on near-Earth observations from][]{Aguilar2015} at large distance.
%while away from the bubble the distribution matches the average Galactic CR one, $f_{\rm gal}$ \cite[tuned on near-Earth observations from][]{Aguilar2015}. 
We can distinguish three regions, labeled from 1 to 3: the cold wind, the shocked wind (i.e. the hot bubble) and the region external to the bubble. In the cold wind the CR distribution can be approximated as
\begin{equation} \label{eqn:f1}
    f_1(r,p) \simeq f_{\rm ts}(p) \, \e^{-v_w (R_{\rm ts}-r)/D_1} \,,
\end{equation}
inside the bubble as:

\begin{equation}
\label{eqn:f2}
    \begin{split}
    f_2(r,p) =& f_{\rm ts}(p) \, \e^{\alpha(r)} \,
        \frac{1 + \beta(\e^{\alpha(R_b)}\e^{-\alpha(r)}-1)}
        {1+\beta(\e^{\alpha(R_b)} -1)} \\
        & + f_\mathrm{gal}(p) \,\frac{\beta\left(\e^{\alpha(r)} - 1\right)}{1+\beta(\e^{\alpha(R_b)} -1)}\, ,
    \end{split}
\end{equation}
while outside as:
\begin{equation}
\label{eqn:f3}
    f_3(r,p) = f_b(p) \frac{R_b}{r} + f_\mathrm{Gal}(p)\left(1-\frac{R_b}{r}\right)
\end{equation}
where $f_b(p) = f_2(R_b,p)$ is the particle distribution at the bubble boundary.
The functions $\alpha$ and $\beta$ are:
\begin{eqnarray}
  \alpha(r,p) &=& \frac{u_2 R_{ts}}{D_2(p)} \left(1-\frac{R_{ts}}{r}\right) \\
  \beta(p) &=& \frac{D_3(p) R_b}{u_2 R_{ts}^2}\, ,
\end{eqnarray}
\noindent
where the diffusion coefficients are assumed to be spatially constant within each of the regions: $D_1$ in the cold wind, $D_2$ in the shocked wind region and $D_3$ in the ISM.
$D_3$ is assumed equal to the average Galactic diffusion coefficient, whereas $D_1$ and $D_2$ are determined by the level of magnetic turbulence inside the bubble, which is largely unknown. For this reason, we explore different types of turbulence spectra (Kolmogorov, Kraichnan and Bohm) normalized in such a way that a fraction $\eta_B= 5\%$ of the wind kinetic energy is converted into magnetic energy, namely 
\begin{equation}
  4\pi r^2 v_w  \frac{\delta B_{w}}{4\pi}  =\eta_B \frac{1}{2} \dot{M} v^2_w \,.    
\end{equation} 
For Kolmogorov and Kraichnan we assume an injection scale fixed to 1~pc (roughly corresponding to the size of the stellar cluster core) while the Bohm case implies that the injection also occurs on smaller scales with similar power. 

The solution at the termination shock is:
\begin{equation}
f_{\rm ts}(p) = \frac{\eta_{\rm CR}}{2} \frac{\sigma n_1 v_w^2}{4 \pi \Lambda (m_p c)^3 c^2}
        \left( \frac{p}{m_p c} \right)^{-\alpha_{\rm CR}}
        \, e^{-\Gamma(p/p_{\max})} \,.
\end{equation}

In the above equation $n_1$ and $v_w$ are the plasma density and speed upstream of the termination shock, while $\sigma$ is the compression ratio at the shock, $m_p$ is the proton mass and $c$ the speed of light and $p_{\max}=E_{\max}/c$ is the maximum momentum of accelerated particles.
Notice that $f_{\rm ts}$ is normalized by requiring that the cosmic-ray luminosity is a fraction of the wind luminosity, namely 
$\eta_{CR} L_w = L_{\rm CR} \equiv 4\pi R_{\rm ts}^2 v_w/\sigma \int E f_{ts}(E) dE$. This determines the normalization constant $\Lambda= \int_{x_{\rm inj}}^{\infty} x^{2-\alpha_{\rm CR}} [(x^2+1)^2-1] \e^{-\Gamma(x)} dx$. 
Finally, the exponential function $e^{-\Gamma}$ depends on the ratio $p/p_{\rm max}$ and on the chosen diffusion coefficient. $E_{\max}$ is derived from the assumptions we made on the magnetic field intensity, and on the diffusion coefficient, equating the diffusion length ($D(E_{max})/V_{w}$) to $R_{TS}$. 
The result is: $E_{\max}=$ 18.8 TeV, 115.3 TeV, and 708.5 TeV in the Kolmogorov, Kraichnan and Bohm regime, respectively. These energies correspond to gamma-rays out of the energy range probed by Fermi. Therefore, $E_{\rm max}$ cannot be constrained using Fermi-LAT data alone. As a consequence, our model depends only on two parameters, the acceleration efficiency {($\eta_{\rm CR}$)} and the spectral slope at injection {($\alpha_{\rm CR}$)}, which we determine by fitting the gamma-ray emission.

\section{Model fitting} \label{app:fit}
\begin{figure*}
    \centering
    \includegraphics[width=1\linewidth]{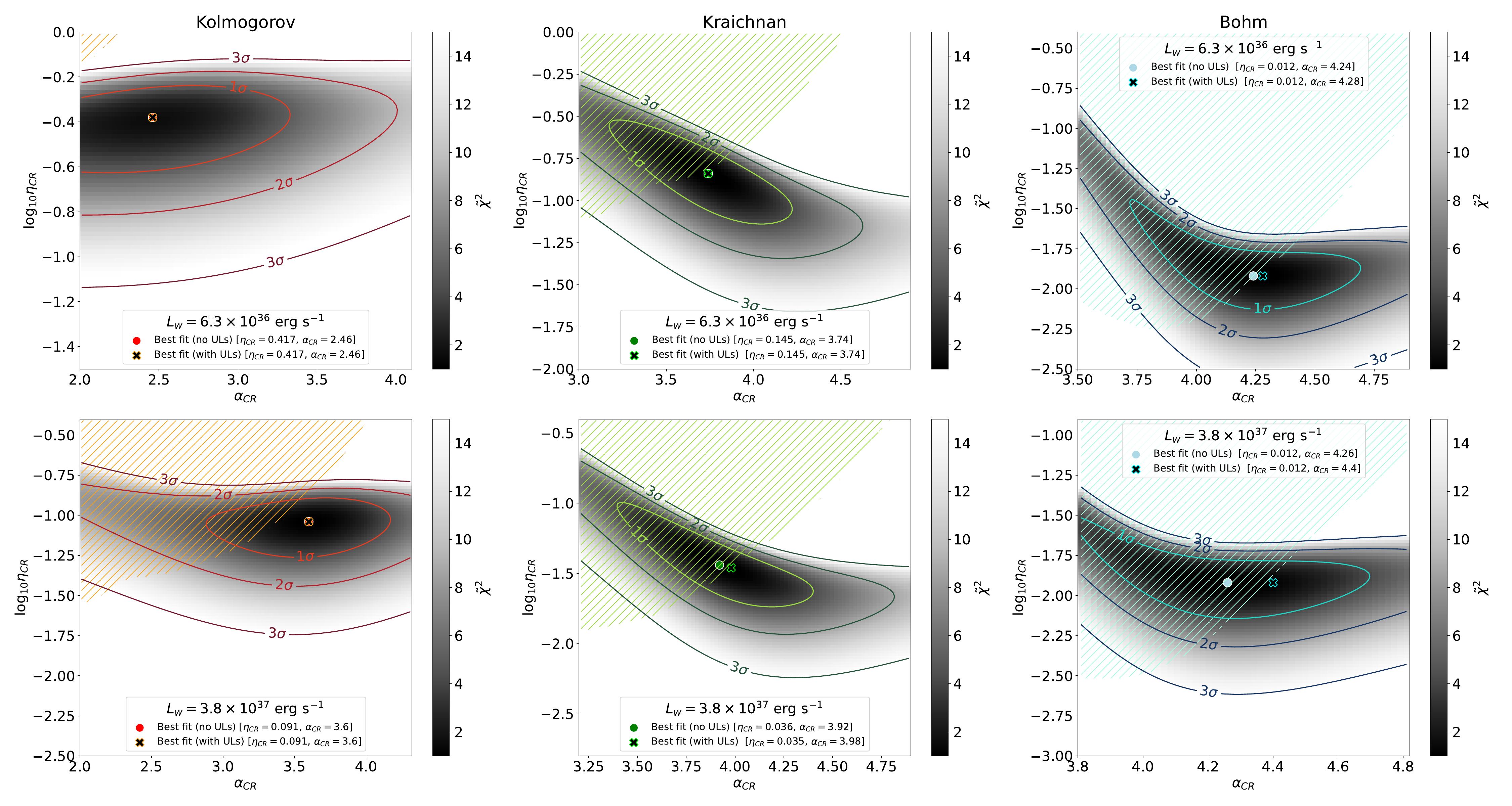}
    \caption{Parameter space explored with the fitting procedure. The darker zones are values with a smaller $\chi^2$, while the shaded regions represent those parameters excluded by the gamma-ray upper limits. The best fit value is indicated as circle, unless it is found in the forbidden parameter region, in which case the new best fit value is indicated as a cross. }
    \label{fig:par_space}
\end{figure*}

\begin{figure*}
    \centering
    \includegraphics[width=0.33\linewidth]{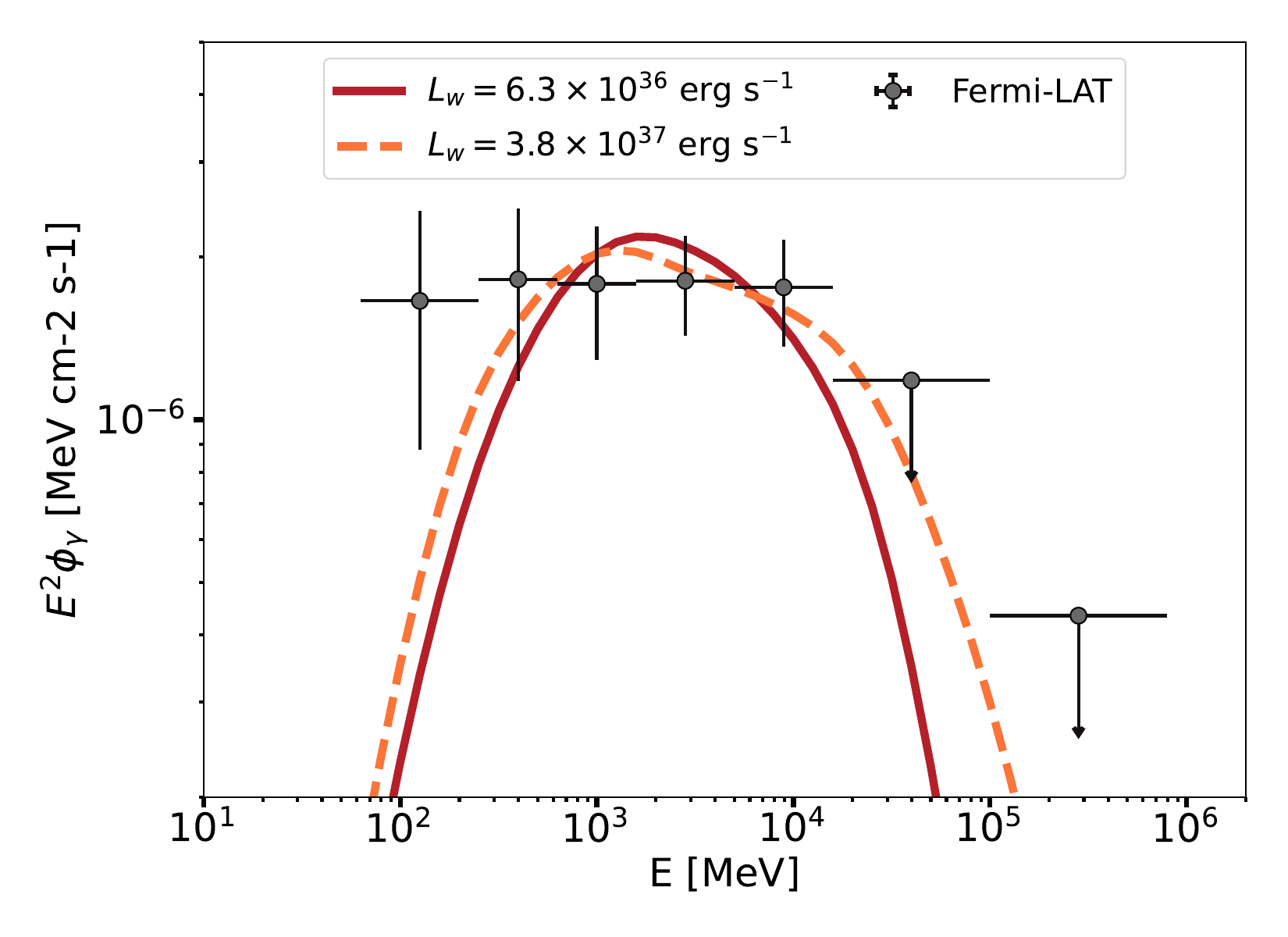}\includegraphics[width=0.33\linewidth]{Figures/BestFit_SEDs_Kra.pdf}\includegraphics[width=0.33\linewidth]{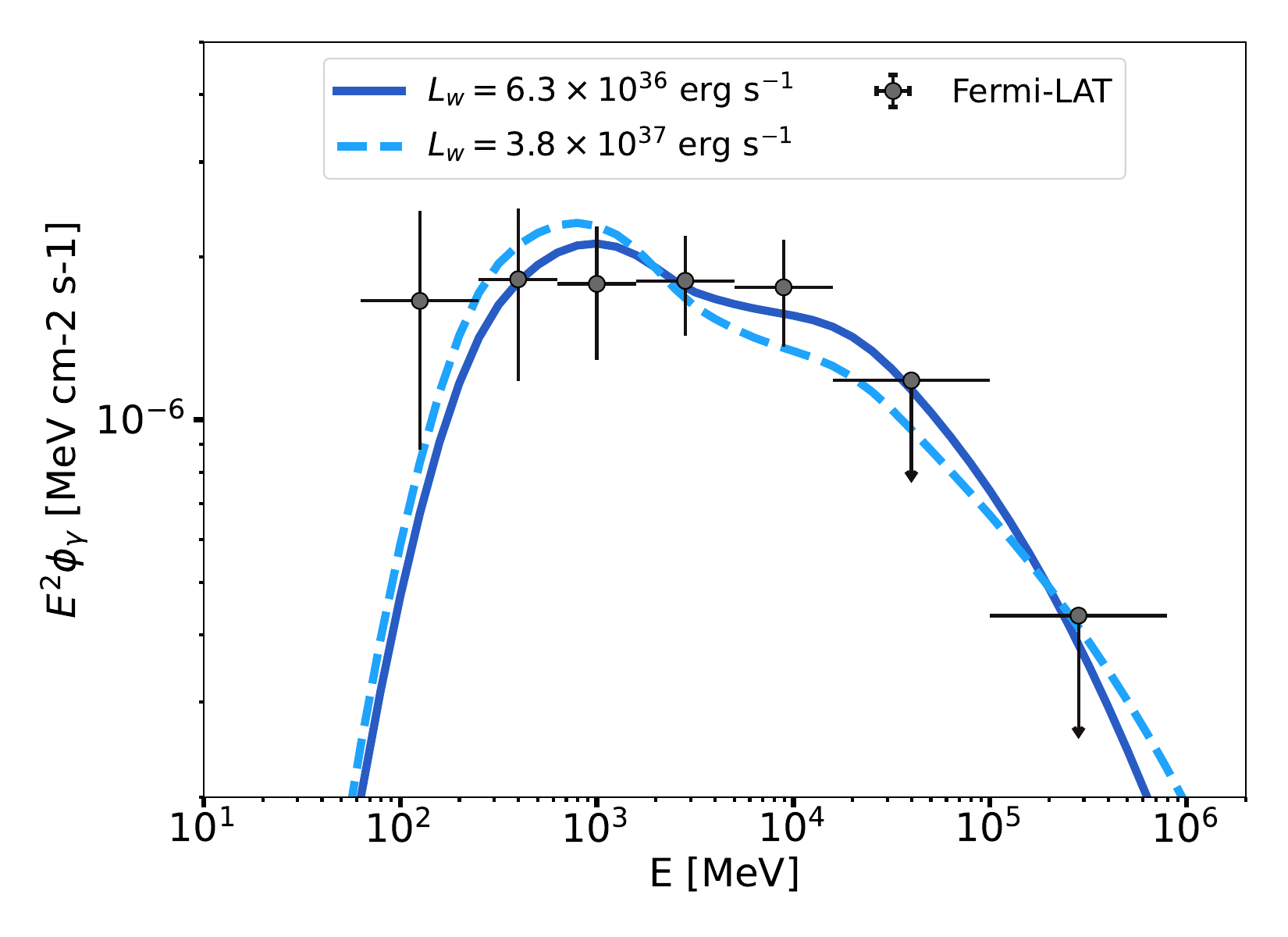}
    \caption{Three different fit results to the spectral energy distribution (SED) of the source detected by Fermi-LAT in correspondence of NGC~6611/M16. The solid and dashed curve differ in the assumed wind luminosity, $L_w$, as indicated in the figure legend. The three different panels refer to the three turbulence spectra assumed: Kolmogorov (red), Kraichnan (green), and Bohm (blue).}
    \label{fig:sed-all}
\end{figure*}

To derive the unconstrained parameters, we fit the model described in Appendix \ref{app:acc_prop} to the derived spectral data, with the caveat of imposing $f_{Gal}(p)=0$, to account that this component is already subtracted in the analysis. 
The two free parameters of our model, namely the acceleration efficiency $\eta_{\rm CR}$ and the injection spectral slope $\alpha_{\rm CR}$, are obtained by a $\chi^2$ fit of the data.  For the spectral index we considered the intervals $\alpha_{CR}^{\rm Kol} \in [2\, (2),4.1\, (4.3)]$, $\alpha_{CR}^{\rm Kra} \in [3\, (3.2),4.9\, (4.9)]$, and $\alpha_{CR}^{\rm Bohm} \in [3.5\, (3.8), 4.9\, (4.8)]$, while for the efficiency we used $\log_{10}\eta_{CR}^{\rm Kol} \in [-1.5\, (-2.5),0\, (-0.3)]$, $\log_{10}\eta_{CR}^{\rm Kra} \in [-2\, (-2.8),0\, (-0.4)]$, and $\log_{10}\eta_{CR}^{\rm Bohm} \in [-2.5\, (-3),-0.4\, (-0.9)]$,
where the pairs correspond to $L_{W,1}$ ($L_{W,2})$. We used equally spaced grids with bin sizes of $\delta\alpha_{\rm CR} = 0.2$ and $\delta \log_{10}\eta_{\rm CR}=0.2$, for the injection spectral slope and the efficiency respectively. 
In addition to the $\chi^2$ evaluation, we further excluded the region of the parameter space that violates
%\removed{We do not account for} 
the upper limits derived from the Fermi-LAT analysis. %\removed{in the $\chi^2$ evaluation, but we exclude the parameter space that violates them} 
The latter appears as a dashed area in Fig. \ref{fig:par_space}, where the distribution of $\chi^2$ is shown for all six cases considered (two values of the wind luminosity times three models for the diffusion coefficient). 
If the best fit set of values fell in the region of parameter space prohibited by the upper limits, we would then choose a second minimum in the allowed area. The corresponding SEDs are shown in \ref{fig:sed-all}. 

\section{The leptonic contribution}
\label{app:lepton}
In this appendix we provide an estimate of a possible leptonic contribution to the gamma-ray emission as due to non-thermal bremmstrahlung and inverse Compton (IC). The former, like pion decay, depends only on the gas density, while IC depends on the photon radiation field density. We considered both the CMB radiation (with a nominal radiation energy density of 0.261 eV cm$^{-3}$) and an infrared radiation field due to the local environment. We inferred the latter from observations, by integrating the best-fit SED derived from the shell region of M16 \citep{Flagey2011}. We found an IR radiation density of $\sim 9$ eV cm$^{-3}$ for the warm dust component at 70~K. 

\begin{figure}[ht!]
\centering
\includegraphics[width=1\linewidth, height=0.75 \linewidth]{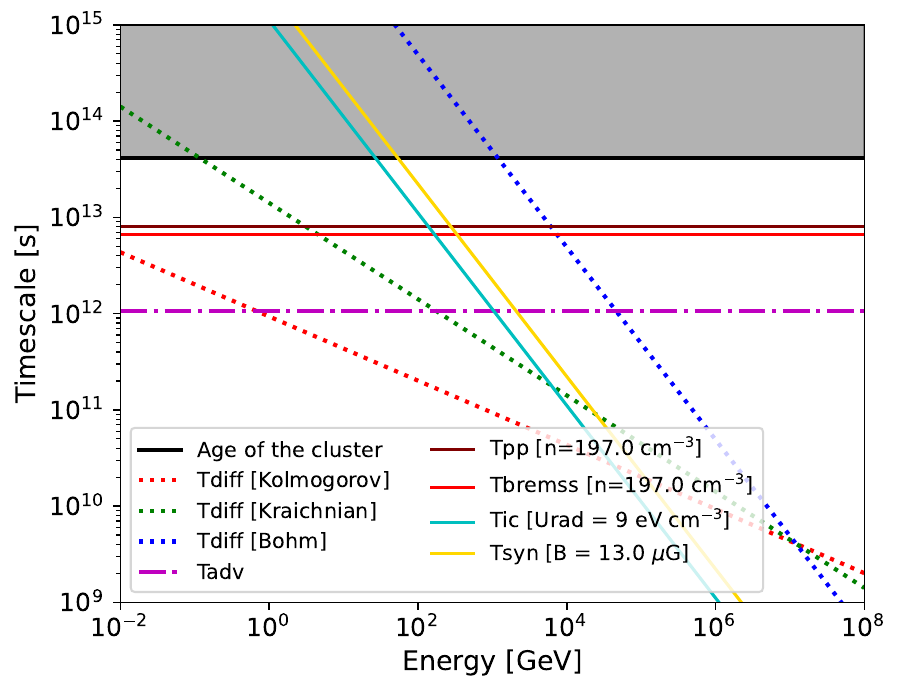}
\includegraphics[width=1 \linewidth, height=0.75 \linewidth]{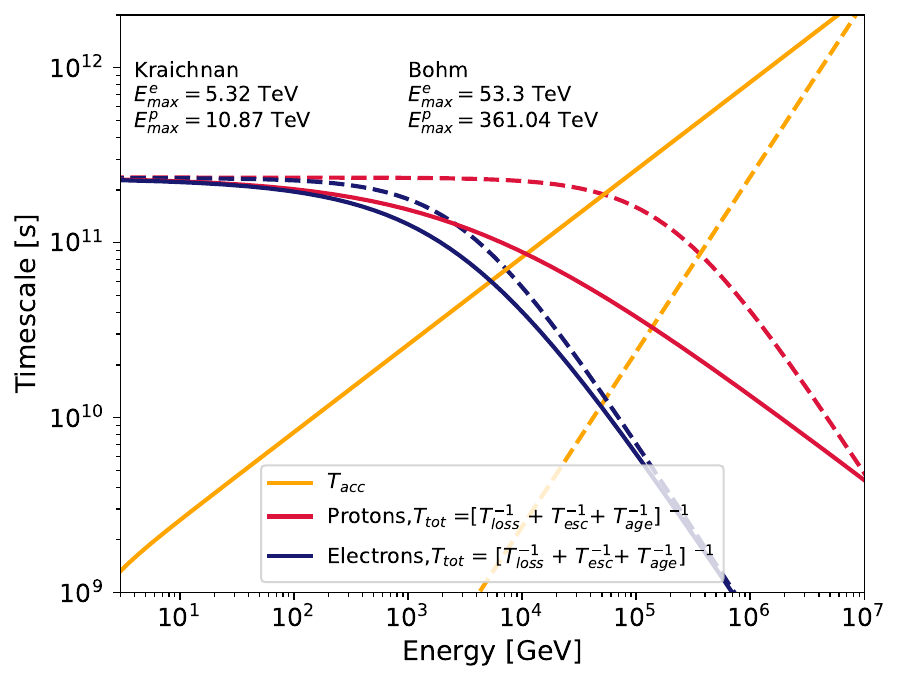}
\caption{The upper panel displays a comparison between the characteristic timescales of all relevant processes inside the bubble: advection, diffusion and energy losses for both protons and electrons. For diffusion, different types of turbulence are considered as detailed in the figure legend. The shaded area represent timescales larger of the age of the system (black line). The lower panel shows the maximum energy of electrons and protons for the Bohm (dashed) and the Kraichnan (solid) cases, obtained as the intersection between the acceleration time and the time-scale that results from the combination of escape ($T_{\rm esc}= (T_{\rm adv}^{-1} + T_{\rm diff}^{-1})^{-1}$), energy losses ($T_{\rm loss}$) and the age of the system $T_{\rm age}$.}.
    \label{fig:timescales}
\end{figure}

As for the electron population, we considered the same injection spectrum as for protons but with a different normalization and a different maximum energy, i.e.
\begin{equation} \label{eq:f_e}
    f_{\rm ts,e}(p) = K_{\rm ep} f_{\rm ts, p}(p) \e^{-p/p_{\max,e}}\ .
\end{equation}
The normalization is scaled to that of protons by a factor $K_{\rm ep}$, left as a free parameter. Notice that typical values of $K_{\rm ep}$ for the case of SNR shocks range between $\sim 10^{-4}$ and few times $10^{-3}$.
The electron maximum energy is determined by comparing the acceleration time with the energy-loss time due to bremsstrahlung, IC and synchrotron emission. For the latter, the assumed magnetic field strength inside the bubble is $\sim$13 $\mu$G, which corresponds to the value estimated downstream of the termination shock assuming $\eta_{\rm B} = 5\%$, as explained in Sec.~\ref{sec:modeling}.
Fig.~\ref{fig:timescales} shows a comparison of all relevant time scales: advection, diffusion, and energy losses for both protons and electrons. The maximum energy can be estimated by comparing the acceleration time scale with the combination of escape and loss times, namely $T_{\rm acc}^{-1}= T_{\rm adv}^{-1}+T_{\rm diff}^{-1}+T_{\rm loss}^{-1}$. This comparison is shown for the cases of Kraichnan and Bohm diffusion in Fig.\ref{fig:timescales}. One can see that the resulting electron maximum energy is a factor $\sim 2$ (7) smaller than for protons in the Kraichnan (Bohm) case. 
To account for the energy losses during propagation inside the bubble, we use the same spatial profile as for protons (Eq.~\eqref{eqn:f1}-\eqref{eqn:f3}) but rescaled by the ratio of the characteristic loss/escape timescale of the two species: 

\begin{equation}
    f_{e}(r,p) = K_{\rm ep} f_{p}(r,p) \, \e^{-p/p_{\max,e}} \times \frac{\left(T_{\rm esc}^{-1} +T_{\rm IC}^{-1} + T_{\rm syn}^{-1} + T_{\rm brem}^{-1}\right)_{e}^{-1}}{\left(T_{\rm esc}^{-1}+T_{\rm pp}^{-1}\right)_{p}^{-1}} \,,
\end{equation}

where $T_{\rm esc}^{-1} = T_{\rm age}^{-1} + T_{\rm adv}^{-1} + T_{\rm diff}^{-1}$.
This approach is an approximation to the real solution, however it is accurate enough for the purpose of this work.

\begin{figure}[ht!]
    \centering
    \includegraphics[width=1\linewidth, height=0.75 \linewidth]{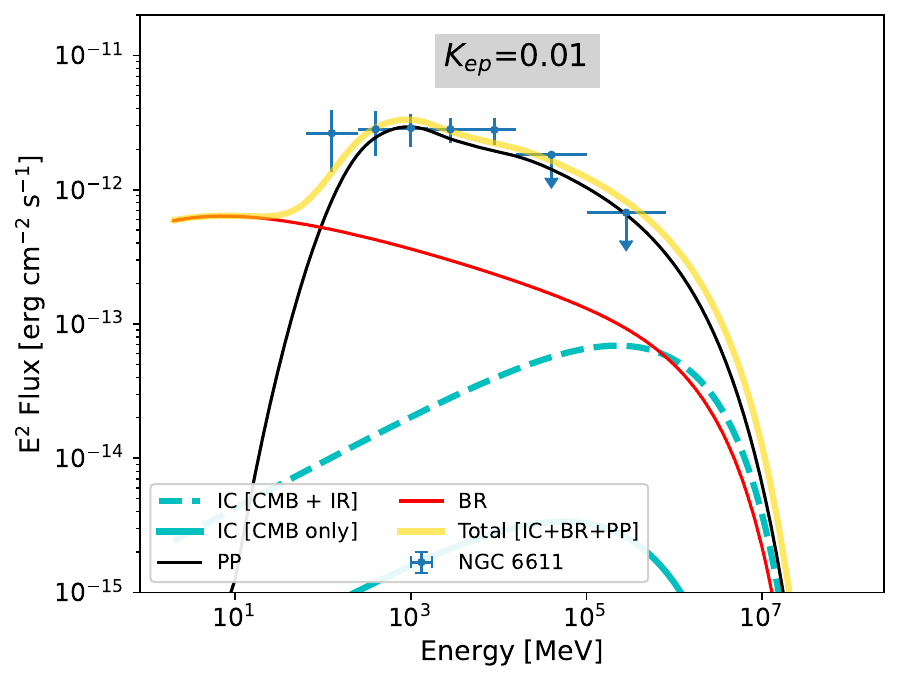}
     \includegraphics[width=1\linewidth, height=0.75 \linewidth]{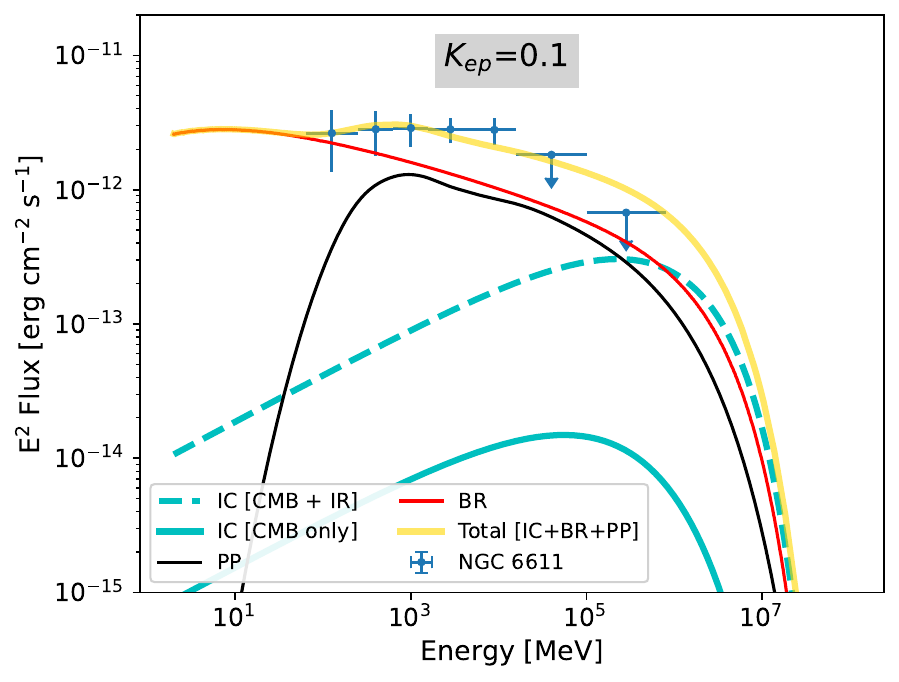}
    \caption{Leptonic contribution to the SED of NGC~6611, considering two different values of the electron-to-proton ratio at injection, $K_{\rm ep}$. The emission is described in terms of bremsstrahlung and inverse Compton. More details on the modeling can be found in the text. The maximum energy of the injected particles is calculated assuming diffusion in Kraichnan-type turbulence. }
    \label{fig:leptonic}
\end{figure}

Once the electron spectrum is determined, we calculate the leptonic emission and we fit the observed gamma-ray data with a combination of Bremsstrahlung, IC and pion decay by changing the value of $K_{\rm ep}$ and $\eta_{\rm CR}$. Two cases for the final SED are reported in Fig. \ref{fig:leptonic} which assume $K_{\rm ep}= 0.01$ and 0.1, and Kraichnan diffusion coefficient.  
The leptonic contribution is dominated by bremmstrahlung in the whole energy range except for the highest energies ($\gtrsim 1$ TeV) where the IC dominates.
It is clear that, in order to have a leptonic contribution comparable to the hadronic contribution without overshooting the data, we need $K_{\rm ep}\gtrsim 0.1$ and, at the same time, a proton acceleration efficiency reduced by a factor $\sim 2$ with respect to the cases shown in Table~\ref{tab:results}. We stress that $K_{\rm ep} \simeq 0.1$ is a very extreme assumption in that much smaller values have been inferred in all cases of shocks where this parameter has been estimated. With more realistic values of $K_{\rm ep}<0.01$ the leptonic contribution is negligible at all energies.

 \end{document}